%% file: sjhiggs.tex
\documentclass[slac_one]{revtex4}
\usepackage{graphicx}
\usepackage{fancyhdr}
\input defs.tex

\begin{document}

\title{Jet substructure as a new Higgs search channel at the LHC}

\author{\underline{Jonathan~M.~Butterworth}, Adam~R.~Davison}
\affiliation{Department of Physics \& Astronomy, University College
  London.}
\author{Mathieu~Rubin, Gavin P.~Salam}
\affiliation{LPTHE; UPMC Univ.\ Paris 6; Univ.\ Paris Diderot; CNRS UMR
  7589; Paris, France.}

\begin{abstract}
  We show that $WH$ and $ZH$ production where the Higgs boson
  decays to $b\bar{b}$ can be recovered as good search channels for the 
  Standard Model Higgs at the Large
  Hadron Collider. This is done by requiring the Higgs to have high transverse momentum, 
  and employing state-of-the-art jet reconstruction and decomposition techniques.
\end{abstract}

\maketitle

\thispagestyle{fancy}

\section{INTRODUCTION} 

A key aim of the Large Hadron Collider (LHC) is to discover 
the Higgs boson, or to prove its non-existence,
and hence elucidate the mechanism of mass generation and electroweak
symmetry breaking. 
Current electroweak fits, together with the LEP exclusion limit,
favour a light Higgs boson, i.e.\ one around 120 GeV in
mass~\cite{Grunewald:2007pm}. This mass region is particularly
challenging for the LHC experiments, and any SM
Higgs-boson discovery is expected to rely on a combination of several
search channels, including gluon fusion $\to H \rightarrow
\gamma\gamma$, vector boson fusion, and associated production with
$t\bar{t}$ pairs~\cite{CSC,cmsphystdr}.

Two significant channels that have generally been considered less
promising are those of Higgs-boson production in association with a
vector boson, $pp\to WH$, $ZH$,
followed by the dominant light Higgs boson decay, to two $b$-tagged
jets.
In this contribution we summarise the work of \cite{Butterworth:2008iy}, 
which presented a way to recover the $WH$ and $ZH$ channels. 

\section{KINEMATIC SELECTION}

Reconstructing $W$ or $Z$ associated $H\to b\bar b$ production would
typically involve identifying a leptonically decaying vector boson,
plus two jets tagged as containing $b$-mesons. However, leptons and 
$b$-jets can be
effectively tagged only if they are reasonably central and of
sufficiently high transverse momentum. The relatively low mass of the
$VH$ (i.e.\ $WH$ or $ZH$) system means that in practice it can be
produced at rapidities somewhat beyond the acceptance, and it is also not
unusual for one or more of the decay products to have too small a
transverse momentum. In addition, there are large backgrounds with intrinsic
scales close to a light Higgs mass. For example, $t \bar t$ events can
produce a leptonically decaying $W$, and in each top-quark rest frame,
the $b$-quark has an energy of $\sim 65$~GeV, a value uncomfortably
close to the $m_H/2$ that comes from a decaying light Higgs boson. If
the second $W$-boson decays along the beam direction, then such a
$t\bar t$ event can be hard to distinguish from a $WH$ signal event.

If one applies kinematic cuts to select $VH$ production in a boosted regime, in
which both bosons have large transverse momenta and are back-to-back,  
the visible cross-section is reduced by a large factor 
(about 20 for $p_T > 200$~GeV). However, the remaining events are those for which
the acceptance of the rest of analysis selection is high. The larger mass of the $VH$ 
system causes it to be central, and the
transversely boosted kinematics of the $V$ and $H$ ensures that their
decay products will have sufficiently large transverse momenta to be
tagged. In addition, the backgrounds are reduced by a larger factor than the signal. Finally, 
the $HZ$ with $Z \to \nu \bar \nu$ channel becomes visible because of the large
missing transverse energy.

In this configuration, the Higgs decay products will be highly collimated, and typically 
found inside a single jet. In the main analysis it was required that this 
Higgs candidate jet should have a $\pt > 200$~GeV.

Three subselections were used for vector bosons: (a) An $e^+e^-$ or
$\mu^+\mu^-$ pair with an invariant mass $80 \GeV < m < 100\GeV$ and
$\pt > \ptmin$.  (b) Missing transverse momentum $> \ptmin$.  (c)
Missing transverse momentum $>30$~GeV plus a lepton ($e$ or $\mu$)
with $\pt > 30$~GeV, consistent with a $W$ of nominal mass with
$\pt > \ptmin$.  

To reject backgrounds we required that there be no leptons with $|\eta|
< 2.5, \pt > 30$~GeV apart from those used to reconstruct the leptonic
vector boson, and no $b$-tagged jets in the range $|\eta|<2.5, \pt >
50$~GeV apart from the Higgs candidate. For channel (c), where
the $t\bar{t}$ background is particularly severe, we require that
there are no additional jets with $|\eta| < 3, \pt > 30$~GeV. 

\section{HIGGS RECONSTRUCTION}

When a fast-moving Higgs boson decays, it produces a single fat jet
containing two $b$ quarks. A successful identification strategy should
flexibly adapt to the fact that the $b\bar b$ angular separation will
vary significantly with the Higgs $p_T$ and decay orientation. In particular
one should capture the $b, \bar b$ and any gluons they emit, while
discarding as much contamination as possible from the underlying event
(UE), in order to maximise resolution on the jet mass. One should
also correlate the momentum structure with the directions of the two
$b$-quarks, and provide a way of placing effective cuts on the $z$
fractions, both of these aspects serving to eliminate backgrounds.
Our method is new, but builds upon prevous work on identifying boosted 
Ws~\cite{Seymour:1993mx,Butterworth:2002tt}.

To flexibly resolve different angular scales we use the inclusive,
longitudinally invariant Cambridge/Aachen (C/A)
algorithm~\cite{Dokshitzer:1997in,Wobisch:1998wt}: one calculates the angular distance
$\Delta R_{ij}^2 = (y_i-y_j)^2 + (\phi_i - \phi_j)^2$ between all
pairs of objects (particles) $i$ and $j$, recombines the closest pair,
updates the set of distances and repeats the procedure until all
objects are separated by a $\Delta R_{ij} > R$, where $R$ is a
parameter of the algorithm. It provides a hierarchical structure for
the clustering, like the \kt algorithm~\cite{Catani:1993hr,Ellis:1993tq},
but in angles rather than in relative transverse momenta (both are
implemented in FastJet~2.3\cite{Cacciari:2005hq}).

Given a hard jet $j$, obtained with some radius $R$, we then use the
following new iterative decomposition procedure to search for a generic
boosted heavy-particle decay. It involves two dimensionless parameters, $\mu$
and $y_\cut$:
\begin{enumerate}
\item Break the jet $j$ into two subjets by undoing its last stage of
  clustering. Label the two subjets $j_1, j_2$ such that $m_{j_1} >
  m_{j_2}$.
\item If there was a significant mass drop (MD), $m_{j_1} < \mu m_{j}$, and
  the splitting is not too asymmetric,  $y= \frac{\min(p_{tj_1}^2,
    p_{tj_2}^2)}{m_j^2} \Delta R_{j_1,j_2}^2 > y_\cut$,
  then deem $j$ to be the heavy-particle neighbourhood and exit the
  loop. 
\item Otherwise redefine $j$ to be equal to $j_1$ and go back to step 1.
\end{enumerate}
The final jet $j$ is the candidate Higgs boson if
both $j_1$ and $j_2$ have $b$ tags. One can then identify $R_{b\bar
  b}$ with $\Delta R_{j_1j_2}$. The effective size of jet $j$ will
thus be just sufficient to contain the QCD radiation from the Higgs
decay, which, because of angular
ordering~\cite{Mueller:1981ex,Ermolaev:1981cm,Bassetto:1984ik}, will
almost entirely be emitted in the two angular cones of size $R_{b\bar
  b}$ around the $b$ quarks.

The two parameters $\mu$ and $y_\cut$ may be chosen independently of
the Higgs mass and $\pt$. Taking $\mu \gtrsim 1/\sqrt{3}$ ensures that if, in
its rest frame, the Higgs decays to a Mercedes $b\bar bg$
configuration, then it will still trigger the mass drop condition (we
actually take $\mu=0.67$).
The cut on $y \simeq \min(z_{j_1},z_{j_2})/\max(z_{j_1},z_{j_2})$
eliminates the asymmetric configurations that most commonly generate
significant jet masses in non-$b$ or single-$b$ jets, due to the soft
gluon divergence. It can be shown that the maximum $S/\sqrt{B}$ for a
Higgs boson compared to mistagged light jets is to be obtained with
$y_\cut \simeq 0.15$. Since we have mixed tagged and mistagged
backgrounds, we use a slightly smaller value, $y_\cut = 0.09$.


A second novel element of our analysis is to \emph{filter} the Higgs neighbourhood. 
This involves rerunning the C/A algorithm with a smaller radius, $R_\filt = \min(0.3, R_{b\bar b}/2)$, and
taking the three hardest objects (subjets) that appear --- thus one
captures the dominant $\order{\as}$ radiation from the Higgs decay, while
contamination from the underlying event. We also require the
two hardest of the subjets to have the $b$ tags.

The results were obtained with
\herwig~6.510\cite{herwig1,herwig2} with {\sc Jimmy}~4.31~\cite{jimmy}
for the underyling event, which has been used throughout the
subsequent analysis. The underlying event
model was chosen in line with the tunes currently used by ATLAS and
CMS (see for example~\cite{hlhc}).  The leading-logarithmic parton shower
approximation used in \herwig~ has been shown to model jet
substructure well in a wide variety of
processes~\cite{Abazov:2001yp,Acosta:2005ix,Chekanov:2004kz,Abbiendi:2004pr,Abbiendi:2003cn,Buskulic:1995sw}. For
this analysis, signal samples of $WH, ZH$ were generated, as well as
$WW,ZW,ZZ,Z+{\rm jet},W+{\rm jet},t\bar{t}$, single top and dijets to
study backgrounds. 

The leading order (LO) estimates of the cross-section were checked by
comparing to next-to-leading order (NLO) results. The $K$-factors were 
such that we do not expect a large effect of the signficance.

\section{RESULTS}

\begin{figure}[t]
  \begin{center}
  \includegraphics[width=0.49\linewidth]{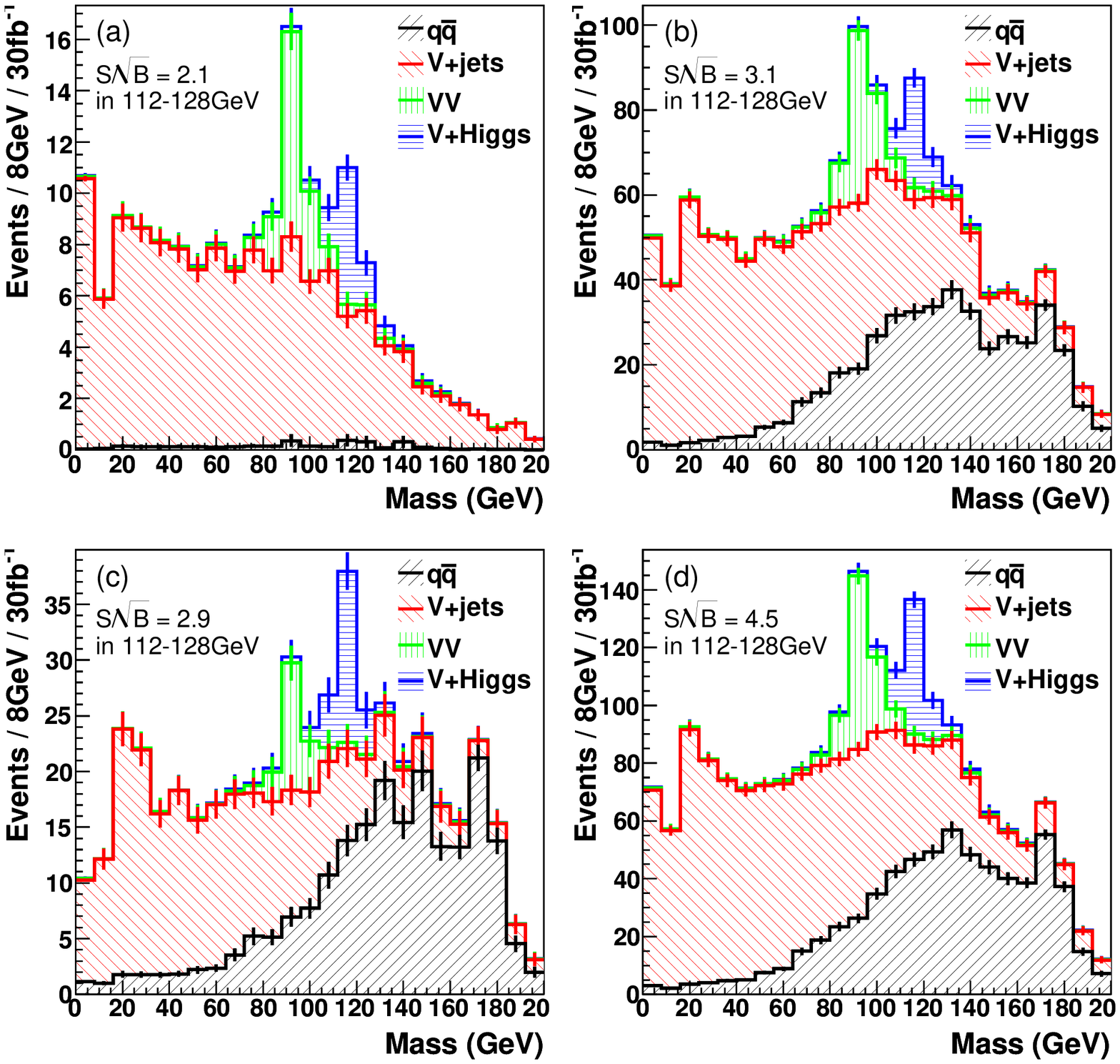}
  \includegraphics[width=0.49\linewidth]{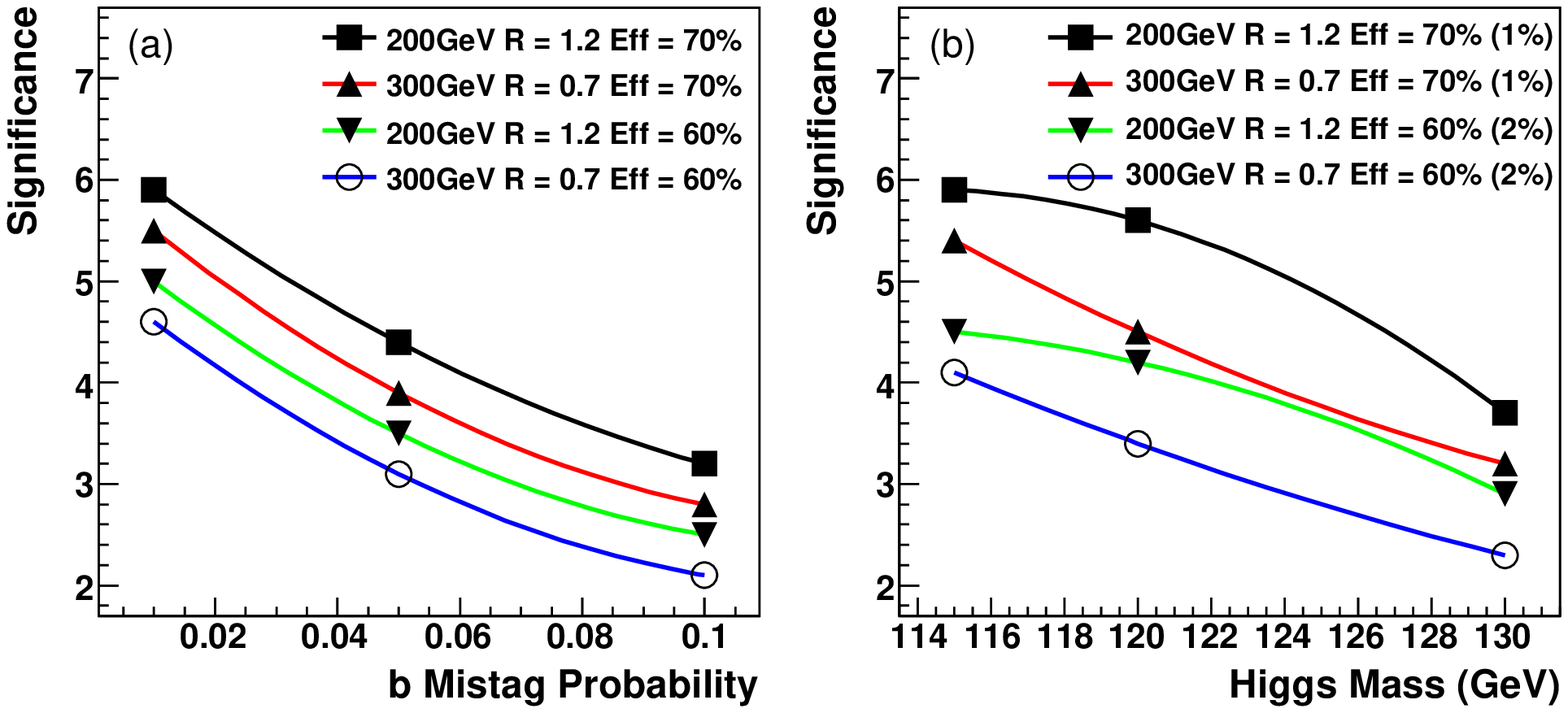}
  \caption{(Left) Signal and background for a 115~GeV SM Higgs simulated
  using \herwig, C/A~MD-F with $R= 1.2$ and $\pt > 200$~GeV, for 30
  fb$^{-1}$. The $b$ tag efficiency is assumed to be 60\% and a mistag
  probability of 2\% is used. The $q\bar{q}$ sample includes dijets
  and $t\bar{t}$. The vector boson selections for (a), (b) and (c) are
  described in the text, and (d) shows the sum of all three
  channels. The errors reflect the statistical uncertainty on the
  simulated samples, and correspond to integrated luminosities $>
  30$~fb$^{-1}$.
  (Right) Estimated sensitivity for 30~fb$^{-1}$ under various
  different sets of cuts and assumptions (a) for $m_H = 115$~GeV as a
  function of the mistag probability for $b$-subjets and (b) as a
  function of Higgs mass for the b-tag efficiency (mistag rates) shown
  in the legend. Significance is estimated as ${\rm signal}/\sqrt{\rm
  background}$ in the peak region.\vspace{-1em}}
  \label{fig:optlep}
  \end{center}
\end{figure}

The results for $R=1.2, \ptmin = 200$~GeV are shown in
Fig.~\ref{fig:optlep}(left), for $m_H = 115$~GeV. The $Z$ peak from $ZZ$ and
$WZ$ events is clearly visible in the background, providing a
critical calibration tool. The major backgrounds are from $W$ or $Z$+jets,
and (except for the $HZ (Z \rightarrow l^+l^-)$ case), $t\bar{t}$. 
Combining the three sub-channels in Fig.~\ref{fig:optlep}d, and
summing signal and background over the two bins in the range
112-128~GeV, the Higgs is seen with a significance of $4.5~\sigma$
($8.2~\sigma$ for 100 fb$^{-1}$). The signal region summed over is 
consistent with the single jet mass resolution for \kt-jets found using 
detailed simulations of the ATLAS detector~\cite{CSC}.

The $b$-tagging and mistag probabilities are critical parameters for
this analysis. Values used by experiments for single-tag
probabilities range up to 70\% for the efficiency and down to 1\% for
mistags. Results for 70\% and 60\% efficiency are summarised in
Fig.~\ref{fig:optlep}a(right) as a function of the mistag probability.

There is a trade-off between rising cross-section and falling fraction
of contained decays (as well as rising backgrounds) as $\ptmin$ is
reduced. As an example of the dependence on this trade-off, we show
the sensitivity for $\ptmin = 300$~GeV, $R = 0.7$ in
Fig.\ref{fig:optlep}a(right).

The significance falls for higher Higgs masses, as shown in
Fig.~\ref{fig:optlep}b(right), but values of $3 \sigma$ or above seem
achievable up to $m_H = 130$~GeV.

\section{Outlook}

Sub-jet techniques have the potential to transform the
high-$\pt$ $WH, ZH (H \rightarrow b\bar{b})$ channel into one of the
best channels for discovery of a low mass Standard Model Higgs at the
LHC.  Realising this potential is a challenge that merits
further experimental study and complementary theoretical investigations.

Jet finding, jet mass and sub-jet technology has come a long way since the previous
round of colliders, and has many applications at the LHC, where we will have interesting 
physics at \cal{O}(100 GeV), and phase space open at \cal{O}(1 TeV). This means that a single
jet often contains 
interesting physics, and it becomes essential to study sub-jet structure. This has already 
been shown for example in applications such as hadronic vector-boson decays from vector-boson 
scattering~\cite{Butterworth:2002tt} and
SUSY decay chains~\cite{Butterworth:2007ke}, and boosted tops \cite{Kaplan:2008ie}, 
including those from from exotic 
resonances~\cite{Holdom:2007nw}. We emphasise that this is a qualitatively new collider 
signature technique at the LHC and has a lot of potential still to be explored.

\bibliographystyle{h-physrev4}

\bibliography{sjhiggs}

\end{document}

%% file: defs.tex
\newcommand{\newc}{\newcommand}

\newcommand{\filt}{\mathrm{filt}}
\newcommand{\order}[1]{{\cal O}\left(#1\right)}
\newcommand{\cut}{\mathrm{cut}}

\newc{\Lumi}{{\cal L}}

\newc{\ra}{\rightarrow}
\newc{\Ra}{\Rightarrow}

\newc{\pom}  {I\hspace{-0.2em}P}

\newc{\gev}{{\rm GeV}}

\newc{\rpv}{{\not \!\! R_p}}
\newc{\rpvm}{{\not  R_p}}

\newc{\gsim}{{\stackrel{>}{\sim}}}
\newc{\lsim}{{\stackrel{<}{\sim}}}
\newc{\sleq} {\raisebox{-.6ex}{${\textstyle\stackrel{<}{\sim}}$}}
\newc{\sgeq} {\raisebox{-.6ex}{${\textstyle\stackrel{>}{\sim}}$}}

\def\3{\ss}

\newc{\ETJ}{E^{{\rm jet}}_T}

\def\ptmin{\hat{p}_T^{\rm min}}

\def\pt{p_T}

\def\kt{K_\perp~}

\def\q2{{\rm Q}^2}
\def\p2{{\rm P}^2}
\def\gev2{{\rm GeV}^{2}}
\def\F2g{F_2^\gamma}
\def\f2c{F_2^{\rm charm}}

\def\d0{D^{0}}

\def\begr{\begin{flushright}}
\def\endr{\end{flushright}}

\def\begl{\begin{flushleft}}
\def\endl{\end{flushleft}}

\def\herwig{HERWIG}

\def\kt{$K_\perp$}

\newcommand{\as}{\alpha_s}

\DeclareMathAlphabet{\mathsc}{OT1}{cmr}{m}{sc}

\newcommand{\GeV}{\,\mathrm{GeV}}